\documentclass[12pt]{article}

\usepackage{amssymb}
\usepackage{amsmath}

\usepackage{graphicx}

\begin{document} %


\title{Probe of a Randall-Sundrum-like model from muon pair production at high energy muon collider}

\author{
S.C. \.{I}nan\thanks{Electronic address: sceminan@cumhuriyet.edu.tr}
\\
{\small Department of Physics, Sivas Cumhuriyet University, 58140,
Sivas, Turkey}
\\
{\small and}
\\
A.V. Kisselev\thanks{Electronic address:
alexandre.kisselev@ihep.ru} \\
{\small A.A. Logunov Institute for High Energy Physics, NRC
``Kurchatov Institute'',}
\\
{\small 142281, Protvino, Russian Federation} }

\date{}

\maketitle

\begin{abstract}
We have examined inclusive $\mu^+\mu^- \rightarrow \mu^+ \mu^- +
E_{\mathrm{miss}}$ and annihilation $\mu^+\mu^- \rightarrow \mu^+
\mu^-$ processes at future high energy muon colliders in the
framework of the Randall-Sundrum-like model with a small curvature
of space-time. The collision energies of 3 TeV, 14 TeV and, 100 TeV
are addressed. Both differential and total cross sections are
calculated, and exclusion bounds on a 5-dimensional gravity scale
are obtained depending on collision energy and integrated luminosity
of the muon colliders.
\end{abstract}

\maketitle


\section{Introduction} %

The Standard Model (SM) has been proven in a lot number of collider
experiments. Nevertheless, we are still searching for solutions for
many problems that SM cannot give a satisfactory solution. One of
such problem is the so-called hierarchy problem which means the
large energy gap between the electroweak scale and gravity scale.
The most elegant answer to this phenomenon has been given in the
framework of the Randall-Sundrum (RS) model \cite{Randall:1999}
which is based on a 5D theory with one extra dimension compactified
in an orbifold $S_1/Z_2$. The main parameters of the RS model are a
compactification radius $r_c$ and AdS$_5$ curvature parameter $k$
(hereinafter referred to as the \emph{curvature} $k$). The model
predicts Kaluza-Klein (KK) gravitons which are heavy resonances with
masses around the TeV scale. The most stringent limits on KK
graviton masses come from the LHC searches for heavy resonances. The
experimental limits depend on a ratio $k/M_\mathrm{Pl}$, where
$M_\mathrm{Pl}$ is the Planck mass. The CMS collaboration have
excluded KK graviton masses below 2.3 to 4.0 TeV for the diphoton
final state \cite{CMS:RS_1}. For the dilepton final state the CMS
have excluded the RS graviton masses in the region 2.47-4.78 TeV
\cite{CMS:RS_2}. The best lower limit of the ATLAS collaboration,
4.6 TeV, has been obtained in searching for the diphoton final state
\cite{ATLAS:RS}.

In papers \cite{Giudice:2005,Kisselev:2005} the RS-like model with a
small curvature of the 5-dimensional space-time (RSSC model) has
been proposed. In particular, a general solution for the warped
metric has been obtained \cite{Kisselev:2016}. In contrast to the
original RS model, the RSSC model has an almost continuous graviton
mass spectrum which is similar to that of the ADD model
\cite{Arkani-Hamed:1998_1}-\cite{Antoniadis:1998}, if $k \ll
M_\mathrm{Pl}$. Thus, the above mentioned experimental bounds are
not applied to the RS scenario with a small value of $k$. A probe of
the RSSC model at the LHC can be found in
\cite{Kisselev:2008,Kisselev:2013}. A detailed comparison of the
RSSC model with the RS model is given in section~2.

In the present paper we intend to examine the RSSC model through the
$\mu^+\mu^- \rightarrow \mu^+ \mu^- + E_{\mathrm{miss}}$ and
$\mu^+\mu^- \rightarrow \mu^+ \mu^-$ processes at a future muon
collider. The idea of the muon collider was proposed by F.~Tikhonin
and G.~Budker in the late 1960's \cite{Tikhonin:1968, Budker:1969},
and it was also discussed in the early 1980's
\cite{Skrinsky:1981,Neuffer:1983}. At present, a great physical
potential of the muon collider for collisions of elementary
particles at very high energies is being actively examined. Its
advantage lies in the fact that muons can be accelerated in a ring
without limitation from synchrotron radiation compared to linear or
circular electron-positron colliders
\cite{Blondel:1999}-\cite{Long:2021}. For instance, the muon
collider may provide a determination of the electroweak couplings of
the Higgs boson which is significantly better than what is
considered attainable at other future colliders
\cite{Barger:1997_2}-\cite{Costantini:2021}. Interest in designing
and building a muon collider is also based on its capability of
probing the physics beyond the SM. In a number of recent papers
searches for SUSY particles \cite{Capdevilla:2021_1}, WIMPs
\cite{Han:2021}-\cite{Franceschini:2022}, and dark matter
\cite{Jueid:2023}, vector boson fusion \cite{Costantini:2020},
leptoquarks \cite{Asadi:2021}, lepton flavor violation
\cite{Bossi:2020}-\cite{Haghighat:2022}, vector-like leptons
\cite{Guo:2023}, heavy leptons \cite{LI:2023,Mekala:2023}, and heavy
neutrinos \cite{Chakraborty:2022}, top Yukawa couplings
\cite{Chen:2022}, multi-boson processes \cite{Bredt:2022}, and
physics of the muon $(g-2)$
\cite{Capdevilla:2021_2}-\cite{Arakawa:2022} are presented. In our
recent paper we have probed axion-like particles (ALPs) at high
energy muon colliders \cite{I_K:ALP}. In a number of papers
anomalous quartic \cite{{Abbot:2022}}-\cite{Yang:2022_2} and triple
\cite{Senol:2022_2,Spor:2022} gauge couplings at the muon collider
were studied. For more details on a spectacular opportunity of the
muon collider in the direct exploration of the energy frontier, see
\cite{MC_report:2022}.

Our goal is to obtain exclusion bounds on the 5-dimensional Planck
scale $M_5$ which can be probed from two mentioned above processes
at TeV and multi-TeV muon colliders. In the inclusive $\mu^+\mu^-
\rightarrow \mu^+ \mu^- + E_{\mathrm{miss}}$ scattering one pair of
muons with large transverse momenta and large invariant mass is
detected, while the other two scattered muons or produced neutrinos
escape a detector. The gravity contribution comes from two
subprocess, $V_1V_2 \rightarrow G \rightarrow \mu^+ \mu^-$, with
$V_{1,2} = \gamma, Z$, and $W^+W^- \rightarrow G \rightarrow \mu^+
\mu^-$, where $G$ denotes a KK graviton, and a summation over all KK
gravitons is assumed. The other process we are interested in is the
annihilation $\mu^+\mu^- \rightarrow \mu^+ \mu^-$ scattering which
has contributions from $s$- and $t$-channel graviton exchanges. Note
that the processes can be easily distinguished experimentally from
each other, since they have quite different distributions in
invariant mass of the detected dimuon pair.

The paper is organized as follows. In the next section, the detailed
description of the RSSC model is presented. In section 3 we examine
the production of the muon pair accompanied by missing energy via
vector boson fusion at the muon collider. The bounds on
5-dimensional Planck scale $M_5$ are obtained. In section 4 we study
the exclusive dimuon production and calculate the values of $M_5$
which can be probed in this collision at the muon collider.

\section{Model of warped extra dimension with small curvature} %

In this section, we describe the RSSC model in detail and compare it
with the original RS model. The RS scenario with one extra dimension
and two branes \cite{Randall:1999} was proposed as an alternative to
the ADD scenario with large flat extra dimensions (EDs)
\cite{Arkani-Hamed:1998_1}-\cite{Antoniadis:1998}. It has the
following background warped metric
\begin{equation}\label{RS_metric}
\quad ds^2 = e^{-2 \sigma (y)} \, \eta_{\mu \nu} \, dx^{\mu} \,
dx^{\nu} - dy^2 \;,
\end{equation}
where $\eta_{\mu\nu}$ is the Minkowski tensor with the signature
$(+,-,-,-)$, $y$ is an extra coordinate, and $\sigma(y)$ is the warp
factor. The periodicity condition $y=y + 2\pi r_c$ is imposed, and
the points $(x_\mu,y)$ and $(x_\mu,-y)$ are identified. Thus, we
have a model of gravity in a slice of the AdS$_5$ space-time
compactified to the orbifold $S^1\!/Z_2$. This orbifold has two
fixed points, $y=0$, and $y=\pi r_c$. There are two branes located at
these points (called Planck and TeV brane, respectively). The SM
fields are confined to the TeV brane.

The classical action of the RS model is given by \cite{Randall:1999}
\begin{align}\label{action}
S &= \int \!\! d^4x \!\! \int_{-\pi r_c}^{\pi r_c} \!\! dy \,
\sqrt{G} \, (2 \bar{M}_5^3 \mathcal{R}
- \Lambda) \nonumber \\
&+ \int \!\! d^4x \sqrt{|g^{(1)}|} \, (\mathcal{L}_1 - \Lambda_1) +
\int \!\! d^4x \sqrt{|g^{(2)}|} \, (\mathcal{L}_2 - \Lambda_2) \;,
\end{align}
where $G_{MN}(x,y)$ is the 5-dimensional metric, with $M,N =
0,1,2,3,4$, $\mu = 0,1,2,3$. The quantities
\begin{equation}
g^{(1)}_{\mu\nu}(x) = G_{\mu\nu}(x, y=0) \;, \quad
g^{(2)}_{\mu\nu}(x) = G_{\mu\nu}(x, y=\pi r_c)
\end{equation}
are induced metrics on the branes, $\mathcal{L}_1$ and
$\mathcal{L}_2$ are brane Lagrangians, $G = \det(G_{MN})$, $g^{(i)}
= \det(g^{(i)}_{\mu\nu})$. The parameter $\bar{M}_5$ is a
\emph{reduced} 5-dimensional Planck scale, $\bar{M}_5 =
M_5/(2\pi)^{1/3}$. The parameter $\Lambda$ is a 5-dimensional
cosmological constant, and $\Lambda_{1,2}$ are brane tensions.

For the first time, the solution for $\sigma(y)$ in
\eqref{RS_metric} has been obtained in \cite{Randall:1999},
\begin{equation}\label{sigma_RS}
\sigma_{\mathrm{RS}}(y) = k |y| \;,
\end{equation}
where $k$ is a parameter with a dimension of mass. It defines the
curvature of the 5-dimensional space-time. Later on in
\cite{Kisselev:2016} the following general solution for $\sigma(y)$
was derived
\begin{equation}\label{sigma}
\sigma(y) = \frac{k r_c}{2} \left[ \left| \mathrm{Arccos} \left(\cos
\frac{y}{r_c} \right) \right| - \left| \pi - \mathrm{Arccos}
\left(\cos \frac{y}{r_c} \right)\right| \right] + \frac{\pi \,|k|
r_c }{2} - C \;,
\end{equation}
where $\mathrm{Arccos(z)}$ is a principal value of the multivalued
inverse trigonometric function $\arccos(z)$, and $C$ is a $y$-independent quantity.%
\footnote{At the same time, $C$ may depend on the parameter $r_c$,
see below.}
The constant $C$ arises in \eqref{sigma} because Einstein--Hilbert's
equations for $\sigma(y)$ contain only $\sigma'(y)$ and
$\sigma''(y)$, where prime denotes a derivative with respect to the
variable $y$, but not the function $\sigma(y)$ itself. Let us
underline that solution \eqref{sigma} (i) obeys the orbifold
symmetry $y \rightarrow - y$; (ii) makes the jumps of $\sigma'(y)$
on both branes; (iii) has explicit symmetry with respect to the
branes. More details can be found in \cite{Kisselev:2016}.

By taking $C=0$ in \eqref{sigma}, we get the RS model
\eqref{sigma_RS}, while taking $C = \pi k r_c$ we come to the
RS-like scenario with the small curvature of space-time (RSSC model,
see \cite{Giudice:2005}-\cite{Kisselev:2016}). In general, different
values of $C$ in the warp function \eqref{sigma} result in different
spectra of the KK gravitons, see Appendix~A for more details.

It is worth to remind the main features of the RSSC model in
comparison with those of the RS model. The interactions of the
Kaluza-Klein (KK) gravitons $h_{\mu\nu}^{(n)}$ with the SM fields on
the TeV brane are given by the effective Lagrangian density
\begin{equation}\label{Lagrangian}
\mathcal{L}_{\mathrm{int}} = - \frac{1}{\bar{M}_{\mathrm{Pl}}} \,
h_{\mu\nu}^{(0)}(x) \, T_{\alpha\beta}(x) \, \eta^{\mu\alpha}
\eta^{\nu\beta} - \frac{1}{\Lambda_\pi} \sum_{n=1}^{\infty}
h_{\mu\nu}^{(n)}(x) \, T_{\alpha\beta}(x) \, \eta^{\mu\alpha}
\eta^{\nu\beta} \;,
\end{equation}
were $\bar{M}_{\mathrm{Pl}} = M_{\mathrm{Pl}}/\sqrt{8\pi}$ is the
reduced Planck mass, $T^{\mu \nu}(x)$ is the energy-momentum tensor
of the SM fields. The coupling constant is equal to
\begin{equation}\label{Lambda_pi}
\Lambda_\pi = \bar{M}_5 \sqrt{\frac{\bar{M}_5}{k}} \;.
\end{equation}
The hierarchy relation looks like
\begin{equation}\label{hierarchy}
\bar{M}_{\mathrm{Pl}}^2 = \frac{\bar{M}_5^3}{k} \left[ e^{2\pi k
r_c} - 1 \right] .
\end{equation}
To compare, in the original RS model the hierarchy relation is
different, $\bar{M}_{\mathrm{Pl}}^2 = (\bar{M}_5^3/k) \left( 1 -
e^{-2\pi k r_c} \right)$, and $\Lambda_\pi = M_{\mathrm{Pl}} e^{-\pi
k r_c}$. As for the $\Lambda$-term and brane tensions,
they obey the fine-tuning relations \cite{Kisselev:2016}%
\footnote{In the RS model $\Lambda = -24 \bar{M}_5^3k^2$, but
$\Lambda_1 = - \,\Lambda_2 = 24 \bar{M}_5^3 k$ \cite{Randall:1999}.}
\begin{equation}\label{fine_tuning}
\Lambda = -24 \bar{M}_5^3k^2 \;, \quad \Lambda_1 = - \,\Lambda_2 =
12 \bar{M}_5^3 k \;.
\end{equation}

In the RSSC model masses of the KK gravitons are proportional to the
curvature $k$ \cite{Giudice:2005,Kisselev:2005},
\begin{equation}\label{graviton_masses}
m_n = x_n k \;, \quad n=1,2, \ldots \;,
\end{equation}
where $x_n$ are zeros of the Bessel function $J_1(x)$. Should we
take $k \ll \bar{M}_5 \sim 1$ TeV, the mass splitting $\Delta m$
will be very small, $\Delta m \simeq \pi k$, and we come to an
\emph{almost continuous} mass spectrum, similar to the mass spectrum
of the ADD model \cite{Arkani-Hamed:1998_1}. Nevertheless, even for
rather small curvature $k$ the RSSC model cannot be regarded as an
IR modification of the ADD model with one ED, see Appendix~B. On a
contrary, in the RS model the gravitons are \emph{heavy resonances}
with masses above one-few TeV. For the first time heavy graviton
searches were discussed in \cite{Davoudiasl:2001}.

As was shown in a number of phenomenological papers on the RSSC
model \cite{Kisselev:2008,Kisselev:2013}, cross sections weakly
depend on the parameter $k$, if $k \ll \bar{M}_5$. That is why, in
what follows we will fix this parameter to be $k = 1$ GeV. Let us
underline once more that the graviton mass spectra are quite
different in the RSSC and RS models. It means that all experimental
bounds and phenomenological predictions for parameters of the
original RS model can not be applied to the RS-like scheme examined
in the present paper.

\section{Production of muon pair with missing energy in muon collisions} %

\begin{figure}[htb]
\begin{center}
\includegraphics[scale=0.5]{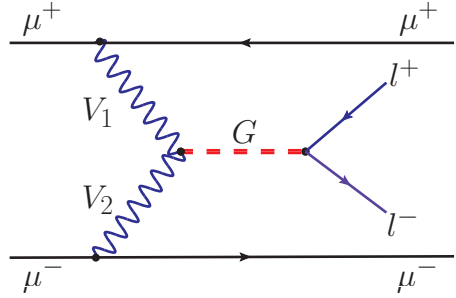}
\caption{The Feynman diagrams describing contribution of the KK
graviton $G$ to the collision of two neutral gauge bosons $V_1, V_2$
with outgoing charged leptons at the muon collider.}
\label{fig:mu-KK-mu}
\end{center}
\end{figure}

\begin{figure}[htb]
\begin{center}
\includegraphics[scale=0.5]{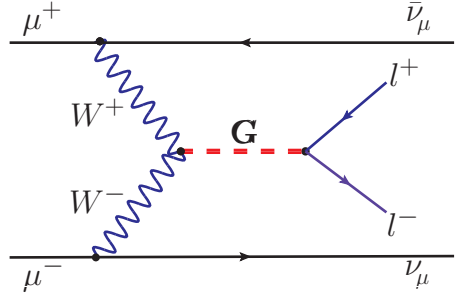}
\caption{The Feynman diagrams describing contribution of the KK
graviton $G$ to the fusion of the $W$ bosons, with two outgoing
charged leptons at the muon collider.}
\label{fig:mu-KK-mu_W}
\end{center}
\end{figure}

Let us study a KK graviton contribution to the inclusive dimuon
production $\mu^{-}\mu^{+} \rightarrow \mu^+ \mu^- +
E_{\mathrm{miss}}$ in the collision of muon beams. It is defined by
gauge boson fusions, $V_1V_2 \rightarrow \mu^+ \mu^-$ ($V_{1,2} =
\gamma, Z$) and $W^+W^- \rightarrow \mu^+ \mu^-$ as shown in
Figs.~\ref{fig:mu-KK-mu} and \ref{fig:mu-KK-mu_W}.%
\footnote{In what follows, $V_1(V_2)$ denotes $\gamma$ or $Z$, while
$V$ without subscript means $\gamma, Z$ or $W$.}
In the framework of the RSSC model a scattering amplitude is equal
to $M_{\mathrm{SM}} + M_{\mathrm{KK}}$, where $M_{\mathrm{SM}}$ is
the SM term, and $M_{\mathrm{KK}}$ is given by a sum of $s$-channel
KK gravitons
\begin{equation}\label{KK_amplitude_in}
M_{KK} = \frac{1}{2\Lambda_{\pi}^2} \sum_{n=1}^{\infty} \left[
\bar{u}(p_{1})\Gamma_{2}^{\mu\nu} v(p_{2}) \,
\frac{B_{\mu\nu\alpha\beta}}{s - m_n^2 + i \, m_n \Gamma_n} \,
\Gamma_{1}^{\alpha\beta\rho\sigma} e_{\rho}(k_{1})e_{\sigma}(k_{2})
\right].
\end{equation}
%
%
Here $k_{1}, k_{2}$, and $e_{\rho}(k_1)$, $e_{\sigma}(k_2)$ are,
respectively, the momenta and polarization vectors of the incoming
bosons, and $p_{1}, p_{2}$ are momenta of the outgoing leptons.
$\Gamma_n$ is the total width of the graviton with the mass $m_n$.
The coherent sum in \eqref{KK_amplitude_in} is over all massive KK
modes. The Feynman rules for the KK graviton were derived in
\cite{Giudice:1999,Han:1999} (see also \cite{Atag:2009}). In
particular, the vector boson-graviton vertex $VVG$ has the form
\begin{equation}\label{Gamma_1}
\Gamma_{1}^{\alpha\beta\rho\sigma} = -\frac{i}{2} \,\{ [m_V^2 +
(k_{1}\cdot k_{2})] \,C^{\alpha\beta\rho\sigma}+
D^{\alpha\beta\rho\sigma} \} \;,
\end{equation}
where $m_V$ is a mass of the gauge boson, and
\begin{align}
C^{\alpha\beta\rho\sigma} &= \eta^{\alpha\rho}\eta^{\beta\sigma}
+\eta^{\alpha\sigma}\eta^{\beta\rho} -
\eta^{\alpha\beta}\eta^{\rho\sigma} \;, \label{tensor_C}
\\
D^{\alpha\beta\rho\sigma} &=\eta^{\alpha\beta}
k^{\sigma}_{1}k^{\rho}_{2} - (\eta^{\alpha\sigma}
k^{\beta}_{1}k^{\rho}_{2} + \eta^{\alpha\rho}
k^{\sigma}_{1}k^{\beta}_{2} - \eta^{\rho\sigma}
k^{\alpha}_{1}k^{\beta}_{2})
\nonumber \\
&- (\eta^{\beta\sigma} k^{\alpha}_{1}k^{\rho}_{2} + \eta^{\beta\rho}
k^{\sigma}_{1}k^{\alpha}_{2} - \eta^{\rho\sigma}
k^{\beta}_{1}k^{\alpha}_{2}) \;.
\label{tensor_D}
\end{align}
The KK graviton-lepton vertex $Gl^+l^-$ is given by
\cite{Giudice:1999,Han:1999}
\begin{equation}\label{Gamma_2}
\Gamma_{2}^{\mu\nu} = -\frac{i}{8} \,
[\gamma^{\mu}(p^{\nu}_{1}-p^{\nu}_{2})+
\gamma^{\nu}(p^{\mu}_{1}-p^{\mu}_{2})] \;.
\end{equation}
Finally, $B_{\mu\nu\alpha\beta}$ in \eqref{KK_amplitude_in} is a
tensor part of the KK graviton propagator. Its explicit expression
was also derived in \cite{Giudice:1999,Han:1999}. We can safely omit
terms in $B_{\mu\nu\alpha\beta}$ which give zero contribution to
eq.~\eqref{KK_amplitude_in}. Then we can write
\begin{eqnarray}\label{graviton_propagator}
B_{\mu\nu\alpha\beta}=\eta_{\mu\alpha}\eta_{\nu\beta}+
\eta_{\mu\beta}\eta_{\nu\alpha}-\frac{2}{3} \,
\eta_{\mu\nu}\eta_{\alpha\beta} \;.
\end{eqnarray}
The $s$-channel contribution of the KK gravitons is equal to
\begin{equation}\label{S_def}
\mathcal{S}(s) =  \frac{1}{\Lambda_{\pi}^2} \sum_{n=1}^{\infty}
\frac{1}{s - m_n^2 + i \, m_n \Gamma_n} \;.
\end{equation}
This series has been analytically calculated in \cite{Kisselev:2006}
\begin{equation}\label{S}
\mathcal{S}(s) = - \frac{1}{4\bar{M}_5^3 \sqrt{s}} \; \frac{\sin
(2A) + i \sinh (2\varepsilon)}{\cos^2 \!A + \sinh^2 \! \varepsilon }
\;,
\end{equation}
where
\begin{equation}\label{A_epsilon}
A = \frac{\sqrt{s}}{k} \;, \quad \varepsilon  = 0.045 \left(
\frac{\sqrt{s}}{\bar{M}_5} \right)^{\!\!3} .
\end{equation}
If the ratio $\sqrt{s}/\bar{M}_5$ is large enough, we get
$\mathcal{S}(s)\simeq - i/(2\bar{M}_5^3 \sqrt{s})$.

The squared amplitude of the subprocess $VV \rightarrow l^-l^+$  is
a sum of three terms,
\begin{equation}\label{M2}
|M|^2 = |M_{\mathrm{SM}}|^{2} + |M_{\mathrm{KK}}|^{2} +
|M_{\mathrm{int}}|^{2}  \;,
\end{equation}
where $M_{\mathrm{SM}}$ denotes the SM amplitude, while
$M_{\mathrm{KK}}$ and $M_{\mathrm{int}}$ are the gravity and
interference terms. In \cite{Atag:2009} the quantities
$|M_{\mathrm{SM}}(\gamma\gamma \rightarrow l^-l^+)|^2$,
$|M_{\mathrm{KK}}(\gamma\gamma \rightarrow l^-l^+)|^2$, and
$|M_{\mathrm{int}}(\gamma\gamma \rightarrow l^-l^+)|^2$ were
calculated for \emph{massless} leptons. The results of our
calculations of squared amplitudes $|M(VV \rightarrow l^-l^+)|^2$
for \emph{nonzero} $m_l$ and $m_V$ are presented in Appendix~C.

The virtual KK graviton production should lead to deviations from SM
predictions in a magnitude of the cross section. The cross section
of the $\mu^{-}\mu^{+} \rightarrow \mu^+ \mu^- + E_{\mathrm{miss}}$
scattering is defined by
\begin{align}\label{cs}
d\sigma &= \int\limits_{\tau_{\min}}^{\tau_{\max}} \!\!d\tau
\!\!\!\int\limits_{x_{\min}}^{x_{\max}} \!\!\frac{dx}{x} \Bigg[
\sum_{V_1, V_2 = \gamma, Z_T,Z_L} \!\!f_{V_1/\mu^+}(x, Q^2)
f_{V_2/\mu^-}(\tau/x, Q^2) \,d\hat{\sigma} (V_1V_2\rightarrow
\mu^+\mu^-) \nonumber \\
&+ \sum_{W_1, W_2 = W_T, W_L} f_{W_1^+/\mu^+}(x, Q^2)
f_{W_2^-/\mu^-}(\tau/x, Q^2) \,d\hat{\sigma} (W_1^+W_2^-\rightarrow
\mu^+\mu^-) \Bigg] ,
\end{align}
where
\begin{equation}\label{y_z_limits}
x_{\max} = 1 - \frac{m_\mu}{E_\mu} \;, \ \tau_{\max} = \left( 1 -
\frac{m_\mu}{E_\mu} \right)^{\!2} , \ x_{\min} = \tau/x_{\max} \;, \
\tau_{\min} = \frac{p_\bot^2}{E_\mu^2} \;,
\end{equation}
and $p_\bot$ is the transverse momenta of the outgoing photons.
$f_{\gamma/{\mu^\pm}}(x, Q^2)$, $f_{Z_T/{\mu^\pm}}(x, Q^2)$,
$f_{Z_L/{\mu^\pm}}(x, Q^2)$, $f_{W_T^\pm/{\mu^\pm}}(x, Q^2)$, and
$f_{W_L^\pm/{\mu^\pm}}(x, Q^2)$ are unpolarized boson distributions
inside unpolarized muon beams. In the leading order they are given
by \cite{Budnev:1975}
\begin{equation}\label{photon_spectrum}
f_{\gamma/\mu^\pm}(x, Q^2) = \frac{\alpha}{2\pi} \frac{1 + (1 -
x)^2}{x} \ln\frac{Q^2}{m_\mu^2} \;,
\end{equation}
and \cite{Lindfors:1987,Ruiz:2021}
\begin{align}\label{massive_boson_spectrum}
f_{Z_T/\mu^\pm}(x, Q^2) &= \frac{\alpha_Z^\pm}{2\pi} \frac{1 + (1 -
x)^2}{x}
\ln\frac{Q^2}{m_Z^2} \;,
\nonumber \\
f_{Z_L/\mu^\pm}(x, Q^2) &= \frac{\alpha_Z^\pm}{\pi} \frac{(1 -
x)}{x} \;,
\nonumber \\
f_{W_T^\pm/\mu^\pm}(x, Q^2) &= \frac{\alpha_W}{2\pi} \frac{1 + (1 -
x)^2}{x} \ln\frac{Q^2}{m_W^2} \;,
\nonumber \\
f_{W_L^\pm/\mu^\pm}(x, Q^2) &= \frac{\alpha_W}{\pi} \frac{(1 -
x)}{x} \;,
\end{align}
where
\begin{equation}\label{alpha_Z}
\alpha_Z^\pm = \frac{\alpha}{(\cos\theta_W \sin\theta_W)^2} \left[
(g_V^\pm)^2 + (g_A^\pm)^2 \right] , \quad \alpha_W =
\frac{\alpha}{4\sin\theta_W^2} \;,
\end{equation}
\begin{equation}\label{gA_gV}
g_V^{\pm} = \pm 1/4 \mp \sin^2\theta_W \;, \quad g_A^{\pm} = \mp 1/4
\;,
\end{equation}
and $m_\mu$ is the muon mass. The variable $x$ in
eqs.~\eqref{photon_spectrum}, \eqref{massive_boson_spectrum} is a
ratio of the boson energy and energy of the incoming muon $E_\mu$.
Note that the $Z$ and $W$ bosons have different distributions for
their transverse ($T$) and longitudinal ($L$) polarizations
\eqref{massive_boson_spectrum}. Note also that the distribution of
the massive bosons V ($V = Z, W$) is suppressed with respect to the
photon distribution by the factor $\ln(Q^2/m_V^2)/\ln(Q^2/m_\mu^2)$.
The mixed $\gamma$--$Z$ and $Z$--$\gamma$ terms are present in
\eqref{cs} along with the $\gamma$--$\gamma$ and $Z$--$Z$ terms.

The differential cross section of the $VV\rightarrow\mu^+\mu^-$
collision is a sum of helicity amplitudes squared,
\begin{equation}\label{subprocess_cs}
\frac{d\hat{\sigma}}{d\Omega} = \frac{1}{64\pi^2 \hat{s}}
\sum_{\lambda_1,\lambda_2,\lambda_3,\lambda_4}
\!\!|M_{\lambda_1\lambda_2\lambda_3\lambda_4}|^2 ,
\end{equation}
where $\sqrt{\hat{s}} = 2E_\mu \sqrt{\tau}$ is an invariant energy
of this collision, and $\lambda_{1,2}$ ($\lambda_{3,4}$) are boson
(muon) helicities. In distributions \eqref{photon_spectrum},
\eqref{massive_boson_spectrum} we put $Q^2 = \hat{s}$.

The scattering angle for high energy muons in
Fig.~\ref{fig:mu-KK-mu} is peaked near $\theta_\mu \approx
0.02^\circ - 1.2^\circ$ \cite{ED_CLIC}. These very forward muons
would most likely escape a muon detector away from colliding beams.
Thus, only muons produced in the fusion of the neutral bosons will
be detected. In the case of the $W$ boson fusion, as in
Fig.~\ref{fig:mu-KK-mu_W}, the neutrino will escape detection.

\begin{figure}[htb]
\begin{center}
\hspace*{-0.4cm}
\includegraphics[scale=0.56]{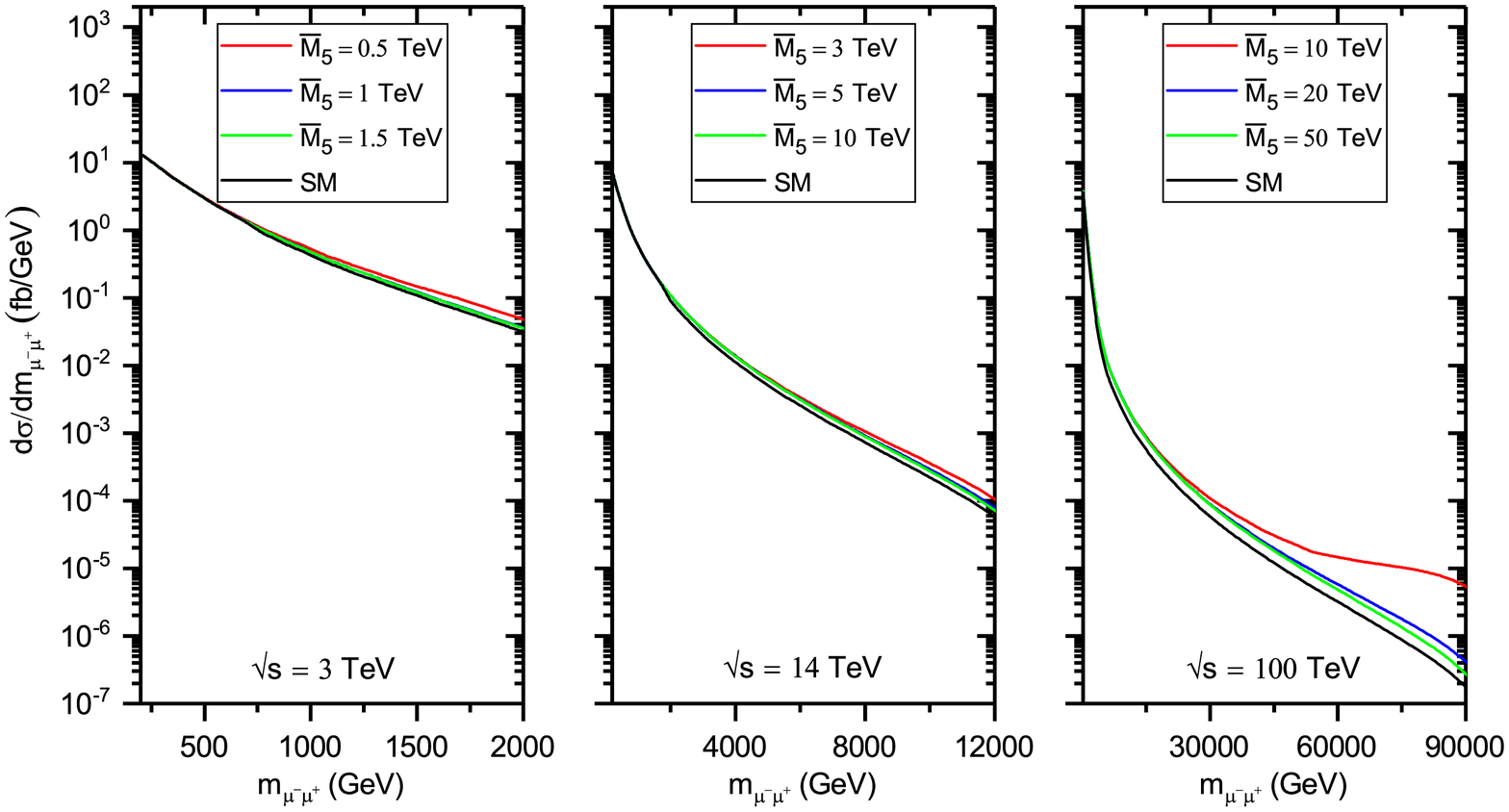}
\caption{The differential cross sections for the process $\mu^+\mu^-
\rightarrow \mu^+ \mu^- + E_{\mathrm{miss}}$ via invariant mass of
two detected muons at the muon collider. The left, middle and right
panels correspond to the colliding energy of 3 TeV, 14 TeV, and 100
TeV, respectively. Three color curves (from the top down) correspond
to different values of $\bar{M}_5$. The SM cross sections (low
curves) are also shown.}
\label{fig:MDCS_N}
\end{center}
\end{figure}

\begin{figure}[htb]
\begin{center}
\hspace*{-0.4cm}
\includegraphics[scale=0.56]{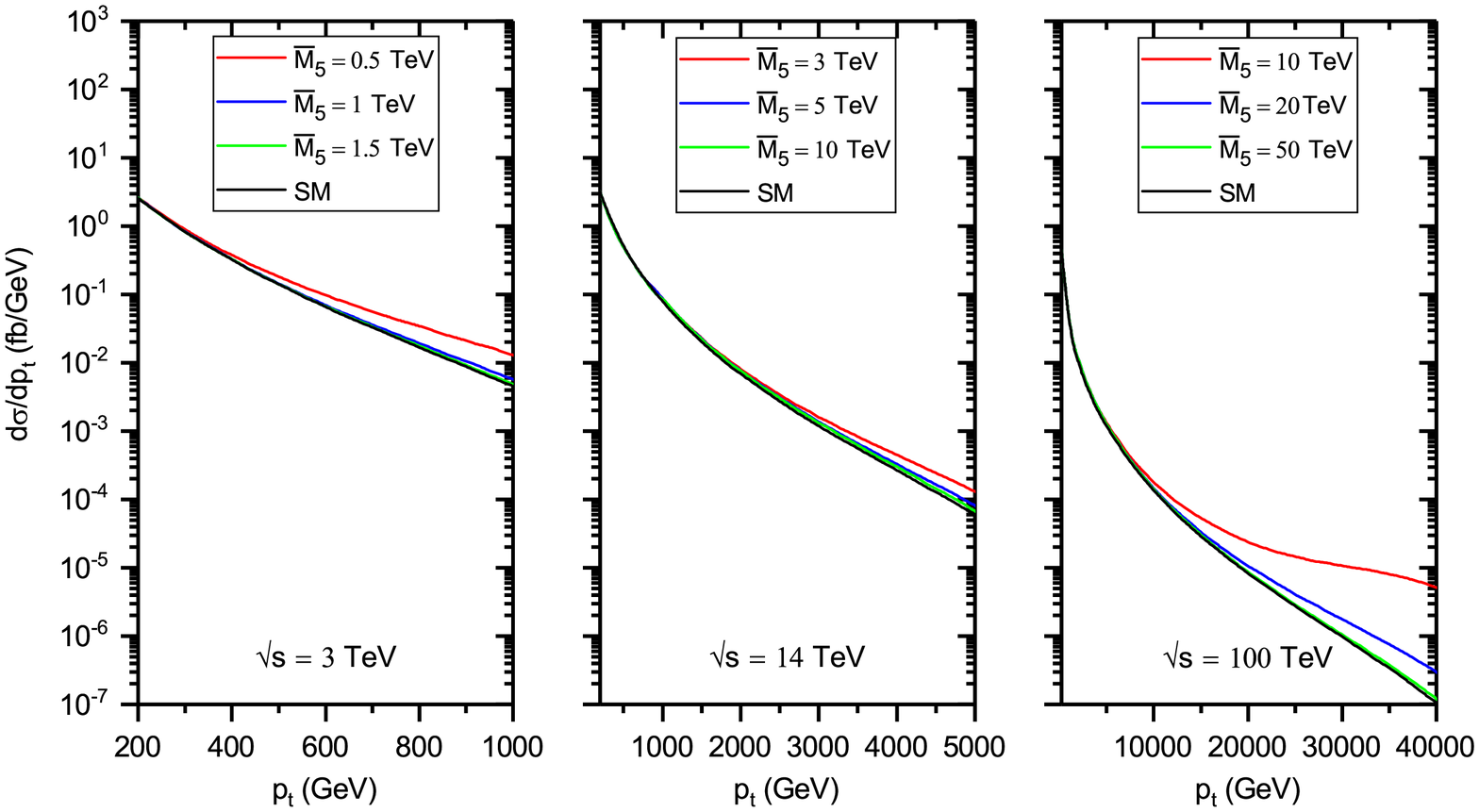}
\caption{The differential cross sections for the process $\mu^+\mu^-
\rightarrow \mu^+ \mu^- + E_{\mathrm{miss}}$ via transverse momentum
of the detected muons at the muon collider.}
\label{fig:PTD2MUMU_N}
\end{center}
\end{figure}

For numerical analysis we apply the cut on the rapidity and
transverse momenta of the detected muons, $|\eta| < 2.5$, $p_t > 50$
GeV. As was already mentioned above, we take $k = 1$ GeV. The
results of our calculations of the differential cross sections for
the $\mu^+\mu^-\rightarrow \mu^+ \mu^- + E_{\mathrm{miss}}$
scattering at the future muon collider are presented in
Fig.~\ref{fig:MDCS_N}. The predictions for three collision invariant
energies of the muon collider are shown. As one can see, for each
energy the cross sections rise as the invariant mass of the detected
muons $m_{\mu^+ \mu^-}$ grows, while the SM cross sections decrease
more rapidly with an increase of $m_{\mu^+ \mu^-}$. We have also
calculated the differential cross sections via transverse momentum
of the detected muons, see Fig.~\ref{fig:PTD2MUMU_N}. They look like
the cross sections in Fig.~\ref{fig:MDCS_N}.

%
\begin{figure}[htb]
\begin{center}
\hspace*{-0.4cm}
\includegraphics[scale=0.56]{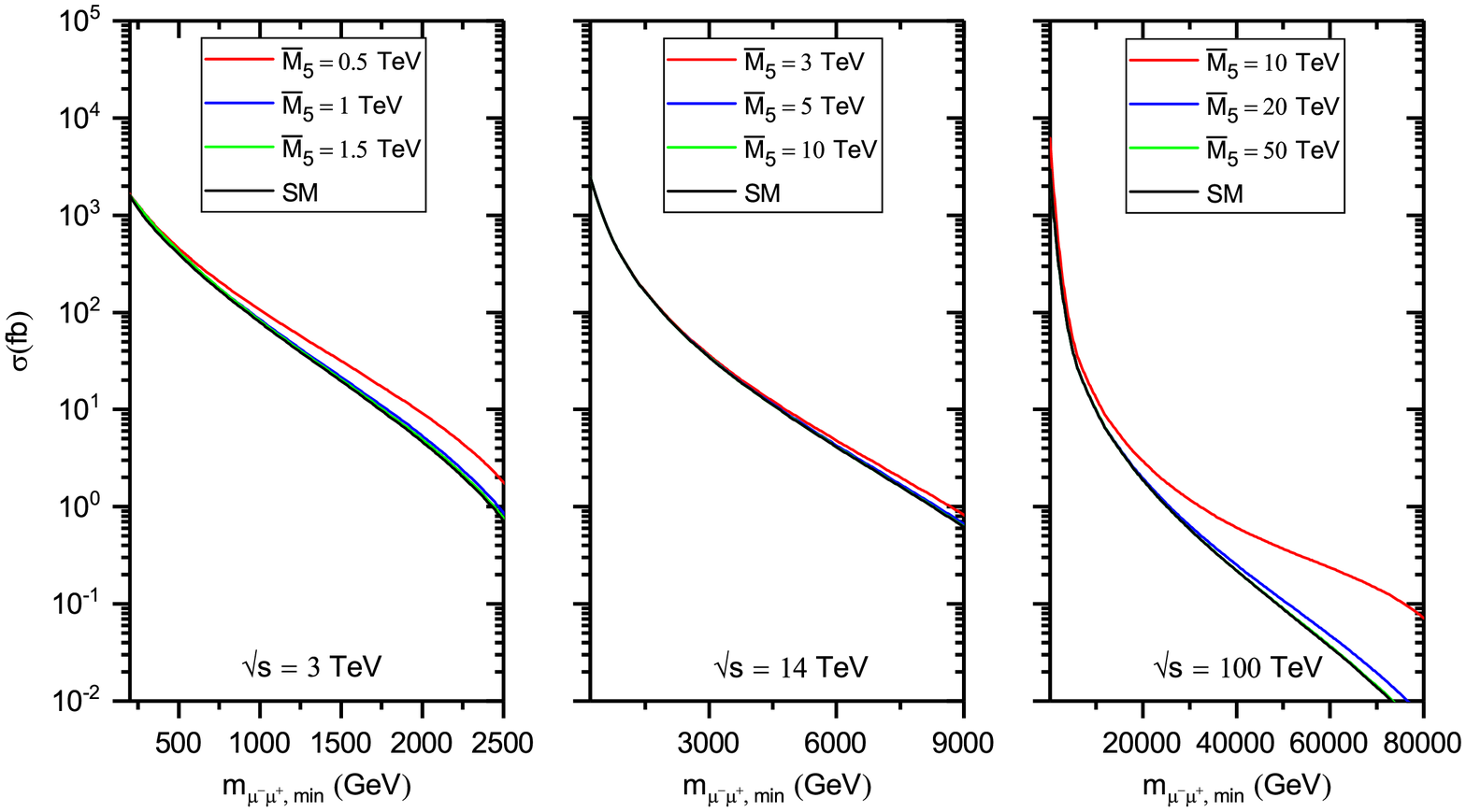}
\caption{The total cross sections for the process $\mu^+\mu^-
\rightarrow \mu^+ \mu^- + E_{\mathrm{miss}}$ via minimal invariant
mass of two detected muons at the muon collider
$m_{\mu^+\mu^-,\min}$.}
\label{fig:MCUTCS_N}
\end{center}
\end{figure}

The total cross section as function of the minimal invariant mass of
two detected muons $m_{\mu^+\mu^-,\min}$ is shown in
Fig.~\ref{fig:MCUTCS_N}. As one can see, it strongly depends on the
gravity scale $\bar{M}_5$. If $\bar{M}_5$ is of order one TeV, the
cross section exceeds the SM one for all collision energies. For
larger values of $\bar{M}_5$ the total cross section strongly
dominates over the SM cross section at $\sqrt{s} = 14$ TeV and
$\sqrt{s} = 100$ TeV.
%

\begin{figure}[htb]
\begin{center}
\hspace*{-0.4cm}
\includegraphics[scale=0.56]{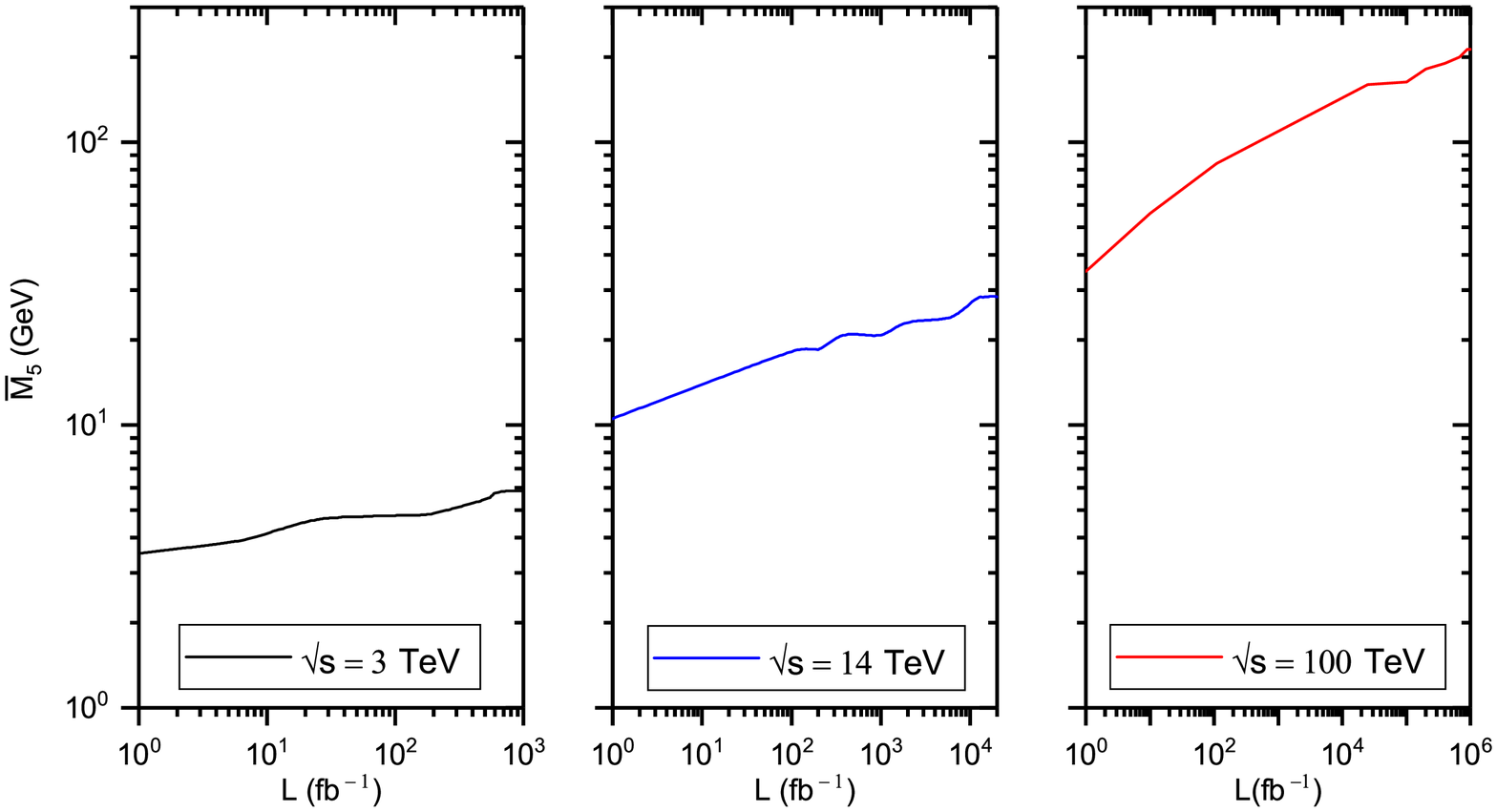}
\caption{The exclusion bounds on the reduced fundamental gravity
scale $\bar{M}_5$ via integrated luminosity of the muon collider for
the process $\mu^+\mu^- \rightarrow \mu^+ \mu^- +
E_{\mathrm{miss}}$. The left, middle and right panels correspond to
the colliding energy of 3 TeV, 14 TeV, and 100 TeV.}
\label{fig:SSM5F_N}
\end{center}
\end{figure}

All this enables us to derive the exclusion bounds on the
5-dimensional reduced Planck scale $\bar{M}_5$. To derive them, we
apply the following formula for the statistical significance $SS$
\cite{SS}
\begin{equation}\label{SS_def}
SS = \sqrt{2[(S - B\,\ln(1 + S/B)]} \;,
\end{equation}
where $S$ is the number of signal events and $B$ is the number of
background (SM) events. We define the regions $SS \leqslant 1.645$
as the regions that can be excluded at the 95\% C.L. To reduce the
SM background, we additionally used the cuts $m_{\mu^-\mu^+}>1$ TeV,
$m_{\mu^-\mu^+}>5$ TeV, and  $m_{\mu^-\mu^+}>50$ TeV for the
collision energy of 3 TeV, 14 TeV, and 100 TeV, respectively. The
results are shown in Fig.~\ref{fig:SSM5F_N}. Our best limits for
$\sqrt{s} = 3$ TeV, $14$ TeV and $100$ TeV are, respectively,
$\bar{M}_5 = 5.86$ TeV, $28.45$ TeV and $213.5$ TeV.

\section{Muon pair production in muon collision} %

In the previous section we assumed that in the $\mu^+\mu^-
\rightarrow \mu^+ \mu^- + E_{\mathrm{miss}}$ process only two final
muons are detected, while two scattered muons or neutrinos escape
the detector. It means that the invariant mass of the outgoing muons
can vary from one event to another. On the contrary, in the
annihilation $\mu^{-}\mu^{+} \rightarrow \mu^+ \mu^-$ process the
invariant mass of the final system $m_{\mu^+\mu^-}$ is fixed and
close to the collision energy $\sqrt{s}$. Note that the muon
collider has a low level of beamstrahlung and synchrotron radiation
compared to linear or circular electron-positron colliders. As a
result, an energy spread in the collision is significantly reduced,
and it enables an improved energy resolution. That is why one can
easily discriminate between two processes by measuring the invariant
mass of the detected muon pair.

\begin{figure}[htb]
\begin{center}
\includegraphics[scale=0.5]{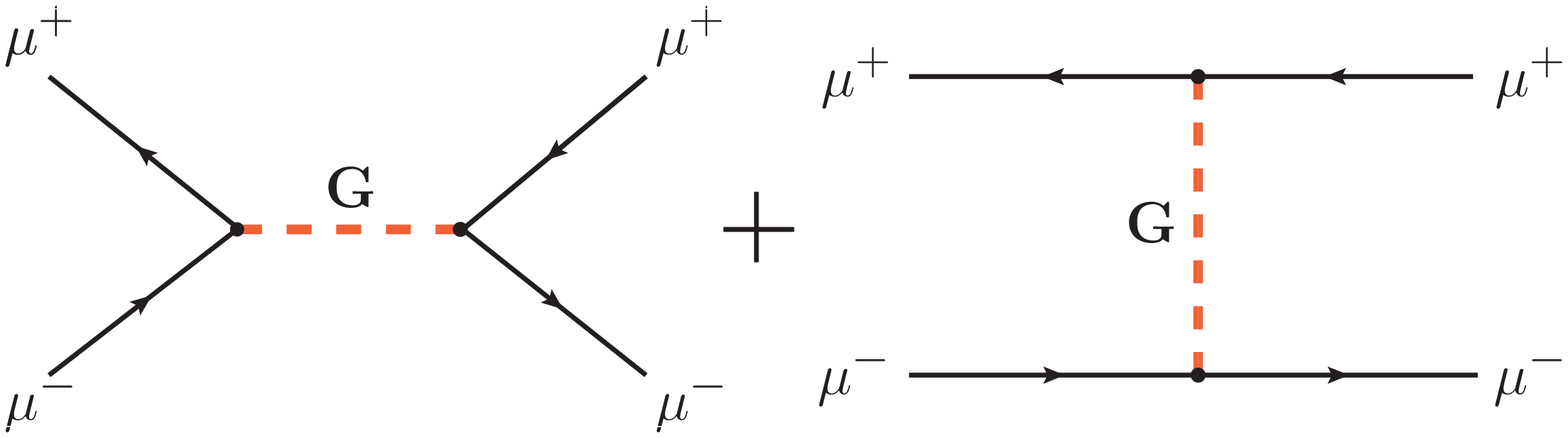}
\caption{The Feynman diagrams describing contribution of the KK
graviton $G$ to the $\mu^{-}\mu^{+} \rightarrow \mu^+ \mu^-$
scattering.}
\label{fig:mu-KK-mu_st}
\end{center}
\end{figure}

\begin{figure}[htb]
\begin{center}
\hspace*{-0.4cm}
\includegraphics[scale=0.56]{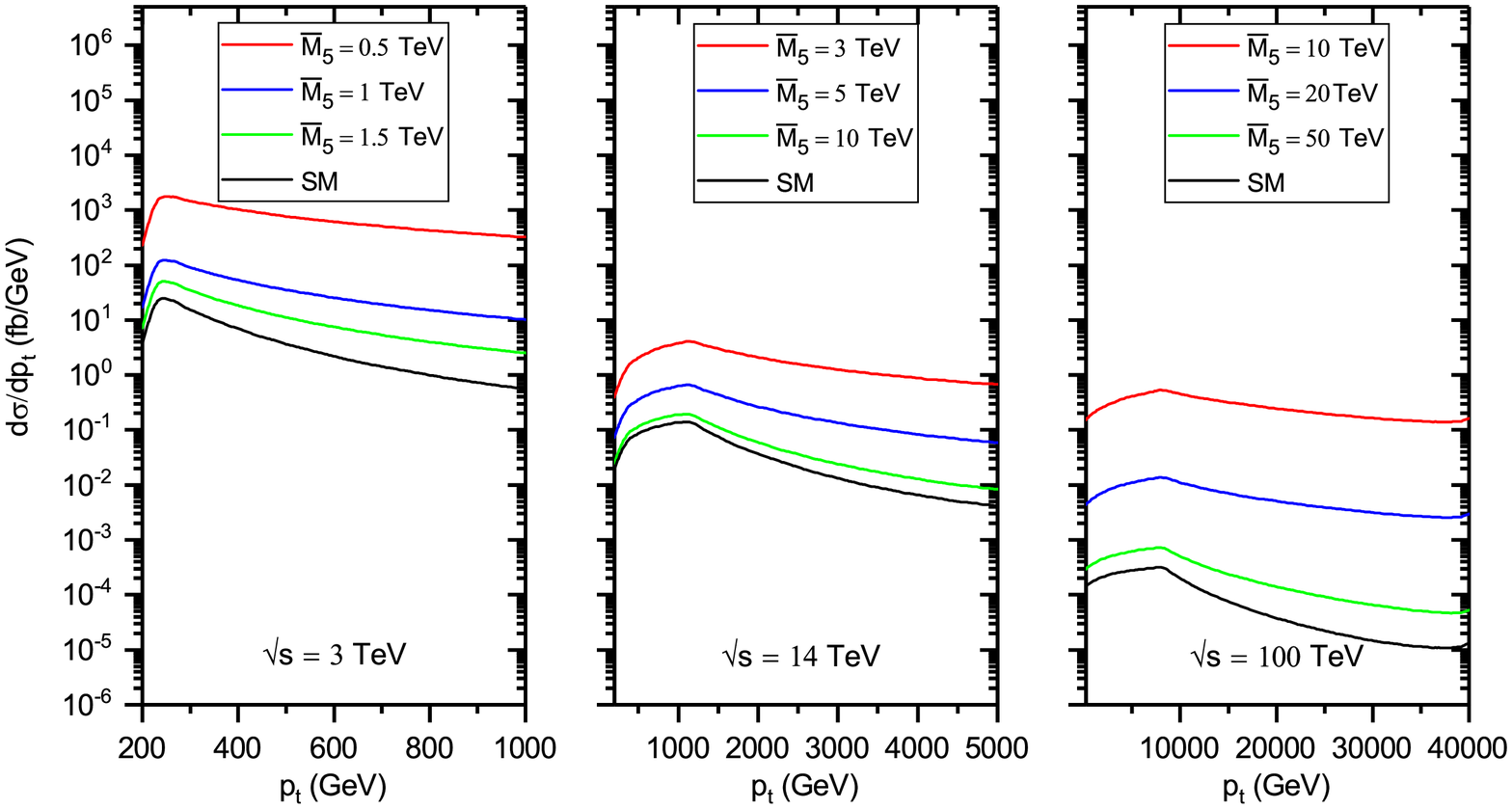}
\caption{The differential cross sections for the process $\mu^+\mu^-
\rightarrow \mu^+ \mu^-$ via transverse momentum of the detected
muons at the muon collider. The left, middle and right panels
correspond to the colliding energy of 3 TeV, 14 TeV, and 100 TeV,
respectively. Three color curves (from the top down) correspond to
different values of $\bar{M}_5$. The SM cross sections (low curves)
are also shown.}
\label{fig:PTDCSMUMU_N}
\end{center}
\end{figure}

\begin{figure}[htb]
\begin{center}
\hspace*{-0.4cm}
\includegraphics[scale=0.56]{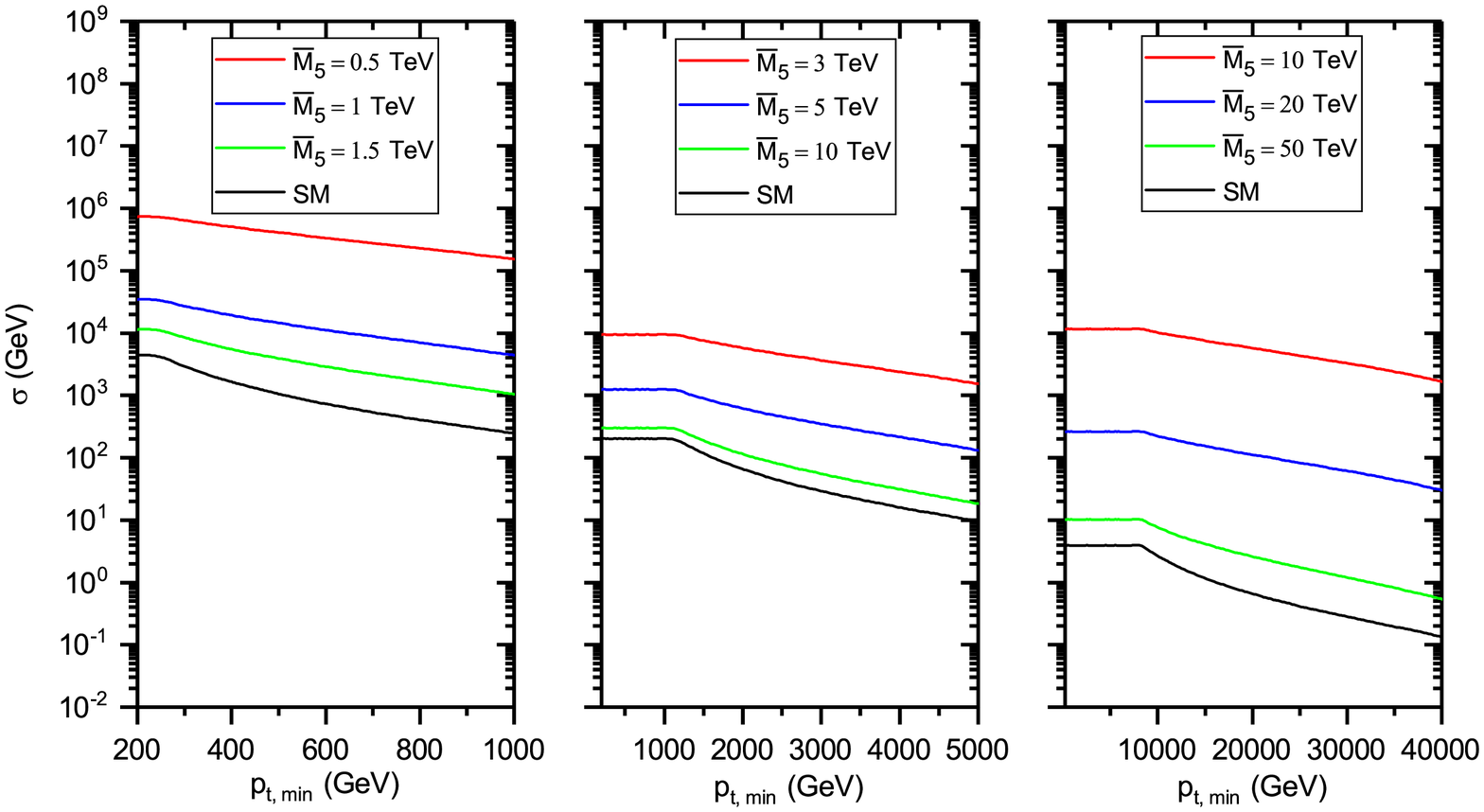}
\caption{The total cross sections for the process $\mu^+\mu^-
\rightarrow \mu^+ \mu^-$ via minimal transverse momentum of the
outgoing muons $p_{t,\min}$.}
\label{fig:PTCUTMUMU_N}
\end{center}
\end{figure}

As in the previous case, a virtual production of the KK gravitons
should give a contribution to the cross sections of the
$\mu^{-}\mu^{+} \rightarrow \mu^+ \mu^-$ scattering. It is shown in
Fig.~\ref{fig:mu-KK-mu_st}.
%
Our analytical expressions for amplitudes squared of this collision
are collected in Appendix~D. It is natural to present the cross
section of the $\mu^{-}\mu^{+} \rightarrow \mu^+ \mu^-$ process as a
function of transverse momentum of the final muons $p_t$. We have
calculated the differential cross sections for the
$\mu^+\mu^-\rightarrow \mu^+ \mu^-$ scattering at the muon collider,
taking into account a contribution from the massive KK gravitons.
The results of our calculations are given in
Fig.~\ref{fig:PTDCSMUMU_N} for three values of the collision energy
$\sqrt{s}$ and different values of the reduced 5-dimensional Planck
scale $\bar{M}_5$. As we can see, for $\sqrt{s} = 14$ TeV and
$\sqrt{s} = 100$ TeV the cross section significantly dominates the
SM one, especially for large $p_t$.
%
The differential cross sections integrated in $p_t$ from the minimal
transverse momentum of the detected muons $p_{t,\min}$ are presented
in Fig.~\ref{fig:PTCUTMUMU_N}.
%

As before, we aim at calculating exclusion bounds on $\bar{M}_5$
which can be probed in the process $\mu^+\mu^- \rightarrow \mu^+
\mu^-$ depending on the integrated luminosity of the future muon
collider. We have used eq.~\eqref{SS_def} for the statistical
significance. In doing so, the cuts $p_{t,\min} = 0.5$ TeV,
$p_{t,\min} = 2.5$ TeV, and $p_{t,\min} = 25$ TeV were applied for
the 3 TeV, 14 TeV, and 100 TeV center-of-mass energies,
respectively.
%
\begin{figure}[htb]
\begin{center}
\hspace*{-0.4cm}
\includegraphics[scale=0.56]{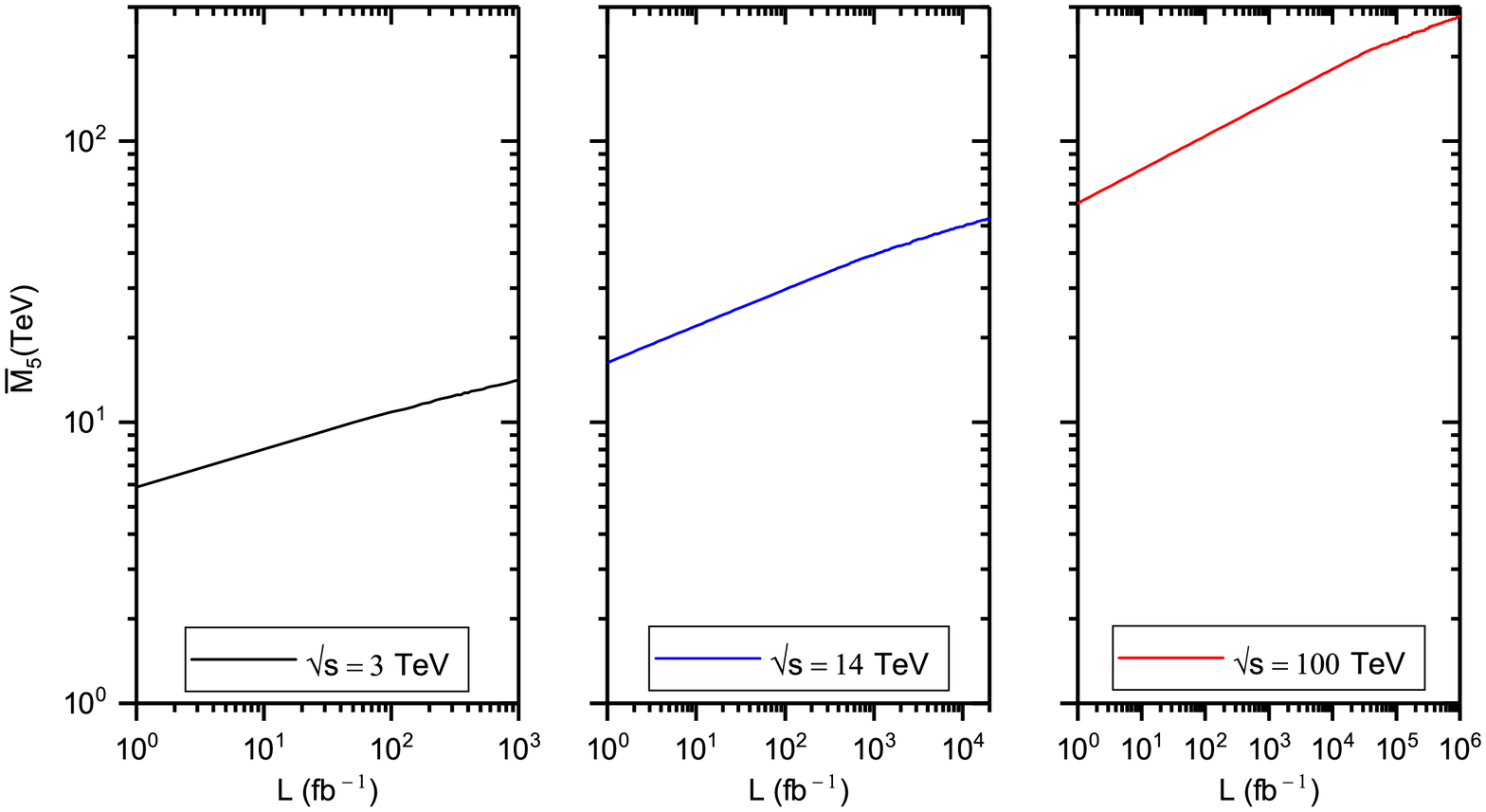}
\caption{The exclusion bounds on the reduced fundamental gravity
scale $\bar{M}_5$ via integrated luminosity of the muon collider for
the process $\mu^+\mu^- \rightarrow \mu^+ \mu^-$. The left, middle
and right panels correspond to the colliding energy of 3 TeV, 14
TeV, and 100 TeV.}
\label{fig:SSM5MUMU}
\end{center}
\end{figure}
The exclusion regions are given in Fig.~\ref{fig:SSM5MUMU}. We
conclude that in the $\mu^+\mu^- \rightarrow \mu^+ \mu^-$ process
scales up to $\bar{M}_5 = 3.85$ TeV, 17.8 TeV and 126.3 TeV can be
probed for $\sqrt{s} = 3$ TeV, $14$ TeV, and $100$ TeV,
respectively. We see that these bounds on $\bar{M}_5$ are stronger
than our constraints for the $\mu^{-}\mu^{+} \rightarrow \mu^+ \mu^-
+ E_{\mathrm{miss}}$ scattering (see section 3).


\section{Conclusions} %

We have examined two collisions at future TeV and multi-TeV muon
colliders in the Randall-Sundrum-like model with the small curvature
(or RSSC model, for short) \cite{Giudice:2005,Kisselev:2005}. It is
the model with one ED and warped metric whose 5-dimensional
space-time curvature $k$ is about one GeV. The other main parameter
of the RSSC model, the reduced 5-dimensional Planck scale
$\bar{M}_5$, is equal to (larger than) one TeV.

We have studied the $\mu^{-}\mu^{+} \rightarrow \mu^+\mu^- +
E_{\mathrm{miss}}$ scattering first. It goes via the $VV \rightarrow
\mu^+\mu^-$ collision, where $V$ means one of the gauge bosons,
$\gamma, Z$, or $W$. The squared amplitudes for the $VV \rightarrow
l^+l^-$ collision, including the gravity, SM and, interference
terms, have been analytically calculated for the first time for
\emph{massive} leptons $l^\pm$. Their explicit expressions are
collected in Appendix~C. The differential cross sections depending
on the invariant mass of the detected muon pair $m_{\mu^+\mu^-}$ are
calculated for three values of the reduced 5-dimensional Planck
scale $\bar{M}_5$ for 3 TeV, 14 TeV and 100 TeV muon colliders. The
total cross sections are presented as functions of the minimal value
of $m_{\mu^+\mu^-}$. As a result, the exclusion bounds on the scale
$\bar{M}_5$ are obtained. They are equal to $\bar{M}_5 = 5.86$ TeV,
$28.45$ TeV and $213.5$ TeV for the collision energy of $\sqrt{s} =
3$ TeV, $14$ TeV and $100$ TeV, respectively.

The $\mu^{-}\mu^{+} \rightarrow \mu^+ \mu^-$ scattering has been
also studied. As in the previous case, we have calculated the
gravity, SM and interference squared amplitudes analytically, see
Appendix~D. It enabled us to estimate numerically the differential
cross sections as functions of the transverse momenta of the
outgoing muons. The total cross sections are also calculated.
Finally, the exclusion bounds on the scale $\bar{M}_5$, have been
obtained. We have shown that the values of $\bar{M}_5 = 3.85$ TeV,
$17.8$ TeV and $126.3$ TeV can be probed at 3 TeV, 14 TeV, and 100
TeV muon colliders. Remember that $\bar{M}_5 = M_5/(2\pi)^{1/3}
\approx 0.54 M_5$, where $M_5$ is the fundamental 5-dimensional
gravity scale $M_5$ in the RSSC model. It means that the bounds on
the scale $M_5$ are approximately twice stronger that the above
mentioned limits for $\bar{M}_5$.

Let us stress again that our bounds on $\bar{M}_5$ should not be
compared with the current experimental limits on the 5-dimensional
gravity scale of the original RS model \cite{Randall:1999}, since
the mass spectra of the KK gravitons and, correspondingly,
experimental signatures for the RSSC and RS models are quite
different \cite{Giudice:2005}-\cite{Kisselev:2016}. The comparative
analysis of these models was presented in section~2.



\setcounter{equation}{0}
\renewcommand{\theequation}{A.\arabic{equation}}

\section*{Appendix A. Shift of warp function}
\label{app:A}

Here we show that a shift of the warp function of the original RS
model,
\begin{equation}\label{sigma_shift}
\sigma(y) \rightarrow \sigma(y) - C \;,
\end{equation}
where $C$ is a constant,  results in a model with a quite different
graviton spectrum. Note that \eqref{sigma_shift} is equivalent to
the rescaling of the 4D metric
\begin{equation}\label{metric_change}
g_{\mu\nu} \rightarrow  g_{\mu\nu} e^{2C} .
\end{equation}
The Einstein tensor $R_{\mu\nu} - (1/2)g_{\mu\nu}R$ is invariant
under such a transformation. As for the energy-momentum tensor, it
is scale-invariant only for \emph{massless} fields. As a simple
example, consider the energy-momentum tensor of the massive scalar
field,
\begin{equation}\label{e-m_tensor}
T_{\mu\nu} = \partial_\mu \phi \, \partial_\nu \phi - \frac{1}{2} \,
g_{\mu\nu} \! \left[ g^{\alpha\beta}  \partial_\alpha \! \phi \,
\partial_\beta \phi  - m^2 \phi^2 \right] .
\end{equation}
It is not scale-invariant due to the third term in
\eqref{e-m_tensor}. In general, theories with \emph{massive} fields
are not invariant under transformation \eqref{metric_change}.

Now consider an effective 4-dimensional gravity action on the TeV
brane (with radion term omitted) \cite{Boos:2002}
\begin{equation}\label{gravity_TeV_action}
S_{\mathrm{eff}} = \frac{1}{4} \sum_{n=0}^{\infty} \int \! d^4x \!
\left[ \partial_\mu h^{(n)}_{\varrho\alpha}(x) \partial_\nu
h^{(n)}_{\delta \lambda} (x) \, \eta^{\mu\nu} - m_n^2
h^{(n)}_{\varrho\alpha} (x) h^{(n)}_{\delta \lambda}(x) \right] \!
\eta^{\varrho\delta} \eta^{\alpha\lambda} \;.
\end{equation}
The transformation \eqref{sigma_shift} can be also regarded as a
rescaling of four-dimensional coordinates
\begin{equation}\label{x_transformation}
x^{\mu} = e^{C} x'^\mu \;,
\end{equation}
but \emph{without change} of the metric \cite{Rubakov:2001}. Note
that \eqref{x_transformation} is not a particular case of general
coordinate transformation in gravity, since the metric tensor
$g_{\mu\nu}$ remains fixed.

The invariance of the action \eqref{gravity_TeV_action} under
transformation \eqref{x_transformation} needs rescaling of the
graviton fields and their mass: $h^{(n)}_{\mu\nu} = e^{-C}
h'^{(n)}_{\mu\nu}$, $m_n = e^{-C} m'_n$. We see that the theory of
massive KK gravitons is not scale-invariant. Only zero mode
(massless graviton) remains unchanged. We conclude that the warp
functions $\sigma(y)$ and $\sigma(y) - C$ correspond to two
\emph{nonequivalent} four-dimensional theories whose spectra of
massive gravitons differ from each other
\cite{Giudice:2005,Kisselev:2016}.



\setcounter{equation}{0}
\renewcommand{\theequation}{B.\arabic{equation}}

\section*{Appendix B. RSSC model versus ADD model with one dimension}
\label{app:B}

We show that the AdS$_5$ space-time differs significantly from a
5-dimensional flat space-time with \emph{one} large ED even for very
small curvature parameter $k$. Compare the hierarchy relations in
both scenarios. The hierarchy relation for the ADD model with one ED
of the size $R_c$ looks like
\cite{Arkani-Hamed:1998_1}-\cite{Antoniadis:1998}
\begin{equation}\label{ADD_hierarchy_relation}
\bar{M}_{Pl}^2 = (2\pi R_c) \, \bar{M}_5^3,
\end{equation}
where $ \bar{M}_5$ is the reduce 5-dimensional Planck scale in the
ADD model. One can see that eq.~\eqref{ADD_hierarchy_relation}
follows from eq.~\eqref{hierarchy} in the limit
\begin{equation}\label{small_k}
\pi k r \ll 1 \; ,
\end{equation}
after replacement $r \rightarrow R_c$. However, it follows from
\eqref{ADD_hierarchy_relation} and \eqref{small_k} that then $k$
should be unnaturally small,
\begin{equation}\label{gravity_scale_over_curvature}
k \ll \frac{\bar{M}_5^3}{\bar{M}_{\mathrm{Pl}}^2} \;,
\end{equation}
even if $\bar{M}_5$ is a few TeV or tens of TeV. That is why, the
RS-like model with the small (but not negligibly small) curvature
$k$ cannot be regarded as an IR modification of the ADD model, at
least, for the parameters used in the present paper.\footnote{Let us
remember that we study the case $k = 1$ GeV, $\bar{M}_5 = 1 \div
100$ TeV.}



\setcounter{equation}{0}
\renewcommand{\theequation}{C.\arabic{equation}}

\section*{Appendix C. Squared amplitudes for $VV \rightarrow l^-l^+$ scattering}
\label{app:C}

Our calculations give the following analytical expressions for the
squared amplitudes of the $\gamma\gamma \rightarrow l^-l^+$
collision in eq.~\eqref{M2}
\begin{align}\label{SM_photons}
|M_{\mathrm{SM}}|^2 &= \frac{8e^4}{(t - m_l^2)^2(s + t - m_l^2)^2}
[-34m_l^8 + m_l^6(60s + 64t)
\nonumber \\
&- m_l^4(31s^2 + 52st + 28t^2) + m_l^2s(s^2 - 2st - 4t^2)
\nonumber \\
&- t(s + t)(s^2 + 2st + 2t^2)] \;,
\end{align}
\begin{align}\label{KK_photons}
|M_{\mathrm{KK}}|^{2} &= -\frac{1}{8}|S(s)|^{2}[ 2m_l^8 - 8m_l^6 t +
m_l^4(s^2 + 4st + 12t^2) - 2m_l^2t(s + 2t)^2
\nonumber \\
&+ t(s + t)(s^2 + 2st + 2t^2) ] \;,
\end{align}
\begin{align}\label{int_photons}
|M_{\mathrm{int}}|^{2} &= -\frac{e^2 [S(s) + S^\star(s)]}{2(t -
m_l^2)(s + t - m_l^2)}[ -2m_l^8 + m_l^6(3s + 4t) + m_l^4s(3s - 4t)
\nonumber \\
&- m_l^2(s^3 + 2s^2t + 3st^2 + 4t^3) + t(s + t)(s^2 + 2st + 2t^2) ]
\;,
\end{align}
where $s$, $t$ are Mandelstam variables, $m_l$ is the lepton mass,
and $S(s)$ is defined in the text \eqref{S_def}-\eqref{A_epsilon}.
If we take $m_l = 0$, we get known results obtained in
\cite{Atag:2009}.

\clearpage

For the $ZZ \rightarrow l^-l^+$ collision our calculations result in
the following formulas
\begin{align}\label{SM_Z_bosons}
|M_{\mathrm{SM}}|^{2}&= \frac{g_Z^{4}}{(t - m_l^2)^2 (s + t - m_l^2
- 2m_Z^2)^2} \{ -2m_l^8[ -300\cos(2\theta_w) + 184\cos(4\theta_w)
\nonumber \\
&- 68\cos(6\theta_w) + 17\cos(8\theta_w) + 195 ]
\nonumber \\
&+ 4m_l^6 [ -4m_Z^2(-94\cos(2\theta_w) + 59\cos(4\theta_w) -
24\cos(6\theta_w)
\nonumber \\
&+ 6\cos(8\theta_w) + 56) + 163s + 196t - 8(33s +
37t)\cos(2\theta_w)
\nonumber \\
&+ 18(9s + 10t)\cos(4\theta_w) + (15s + 16t)(\cos(8\theta_w)-
4\cos(6\theta_w)) ]
\nonumber \\
&+ m_l^4[ -2m_Z^4( -164\cos(2\theta_w) + 128\cos(4\theta_w)
\nonumber \\
&+ 23(-4\cos(6\theta_w) + \cos(8\theta_w) + 3) )
\nonumber \\
&+ 8m_Z^2( 72s + 65t - 4(32s + 27t)\cos(2\theta_w) + (84s +
68t)\cos(4\theta_w)
\nonumber \\
&+ (10s + 7t)(\cos(8\theta_w) - 4\cos(6\theta_w)) ) - 301s^2 -
436t^2 - 668st
\nonumber \\
&+ 4(125s^2 + 260st + 156t^2)\cos(2\theta_w) - 8(39s^2 + 78st +
46t^2)\cos(4\theta_w)
\nonumber \\
&- (31s^2 + 52st + 28t^2)(\cos(8\theta_w) - 4\cos(6\theta_w)) ]
\nonumber \\
&+ m_l^2[ 7s^3 + 50s^2t + 92st^2 + s(s^2 - 2st -
4t^2)(\cos(8\theta_w) - 4\cos(6\theta_w))
\nonumber \\
&+ 80t^3 - 4(3s^3 + 14s^2t + 24st^2 + 24t^3)\cos(2\theta_w)
\nonumber \\
&+ 8(s + 2t)(s^2 + st + 3t^2)\cos(4\theta_w) + 2m_Z^2( 79s^2 + 252st
+ 112t^2
\nonumber \\
&- 4(31s^2 + 108st + 52t^2)\cos(2\theta_w) + 8(9s^2 + 34st +
17t^2)\cos(4\theta_w)
\nonumber \\
&+ (5s^2 + 28st + 16t^2)(\cos(8\theta_w) - 4\cos(6\theta_w)) )
\nonumber \\
&+ 2m_Z^4( - 263s - 414t + (428s + 696t)\cos(2\theta_w) - 16(16s +
27t)\cos(4\theta_w)
\nonumber \\
&- 21(s + 2t)(\cos(8\theta_w) - 4\cos(6\theta_w)) )
\nonumber \\
&+ 2m_Z^6(-156\cos(2\theta_w) + 96\cos(4\theta_w) -
36\cos(6\theta_w) + 9\cos(8\theta_w) + 91) ]
\nonumber \\
&- [ -28\cos(2\theta_w) + 16\cos(4\theta_w) - 4\cos(6\theta_w) +
\cos(8\theta_w) + 19 ]
\nonumber \\
&\times [ 4m_Z^8 - 4m_Z^6(s+3t) + m_Z^4(s^2 + 6st + 14t^2) - 2m_Z^2
t(s + 2t)^2
\nonumber \\
&+ t(s + t)(s^2 + 2st + 2t^2) ] \} \;,
\end{align}
\begin{align}\label{KK_Z_bosons}
|M_{\mathrm{KK}}|^{2}&= \frac{1}{288} |S(s)|^2 \{ -72m_Z^8 +
6m_Z^6(-40m_l^2 + 9s + 48t)
\nonumber \\
&- 4m_Z^4[ -2m_l^2(5s + 96t) + 136m_l^4 + 9t(7s + 12t) ]
\nonumber \\
&+ 3m_Z^2[ - 80m_l^6 + m_l^4(256t - 14s) - 4m_l^2(s^2 + 29st +
68t^2)
\nonumber \\
&+ 9s^3 + 42s^2t + 114st^2 + 96t^3) ] - 36[ m_l^4 - 2m_l^2 t + t(s +
t) ]
\nonumber \\
&\times ( 2m_l^4 -4m_l^2t + s^2 + 2st + 2t^2 ) \} \;,
\end{align}
\begin{align}\label{int_Z_bosons}
|M_{\mathrm{int}}|^{2} &= -\frac{g_Z^{2}[S(s) + S^\star(s)]}{96(t -
m_l^2)(s + t - m_l^2 - 2m_Z^2)} \{ - 12m_l^8[ \cos(4\theta_w) -
2\cos(2\theta_w) ]
\nonumber \\
&+ 2m_l^6[ (16m_Z^2 + 3(3s + 4t)(\cos(4\theta_w) - 2\cos(2\theta_w))
+ 6(s - 2t) ]
\nonumber \\
&+ 2m^4_l[ - m^2_Z((9s + 4t)(\cos(4\theta_w) - 2\cos(2\theta_w))
\nonumber \\
&+ 24s + 60t) + 8m^4_Z(-2\cos(2\theta_w) + \cos(4\theta_w) + 4) +
6(3s^2 + st + 6t^2)
\nonumber \\
&+ 3s(3s - 4t)(\cos(4\theta_w) - 2\cos(2\theta_w)) ]
\nonumber \\
&+ m^2_l[ 4m_Z^2( (2s^2 - 3st + 14t^2)(\cos(4\theta_w) -
2\cos(2\theta_w))
\nonumber \\
&+ 2s^2 + 5st + 46t^2 ) + 4m_Z^4((s - 10t)(\cos(4\theta_w) -
2\cos(2\theta_w)) - 38t)
\nonumber \\
&+ 8m_Z^6( -2\cos(2\theta_w) + \cos(4\theta_w) + 5 )
\nonumber \\
&- 3( 3s^3 + 10s^2t + 2(s^3 + 2s^2t + 3st^2 + 4t^3)
\nonumber \\
&\times (\cos(4\theta_w) - 2\cos(2\theta_w)) + 24st(s + t) ) ]
\nonumber \\
&+ 6[ -2\cos(2\theta_w) + \cos(4\theta_w) + 2 ] [ 2m_Z^8 + m_Z^6(s -
8t)
\nonumber \\
&+ m_Z^4(-s^2 + 2st + 12t^2) - m_Z^2t(s^2 + 7st + 8t^2)
\nonumber \\
&+ t(s + t)(s^2 + 2st + 2t^2) ] \} \;,
\end{align}
where $\theta_w$  is the Weinberg angle, $g_Z =
e/[\sin(\theta_w)\cos (\theta_w)]$, and $m_Z$ is the mass of the $Z$
boson. Because of conservation of helicity, in the massless limit
$m_l = m_Z = 0$ $s$-channel graviton amplitudes squared
\eqref{KK_photons} and \eqref{KK_Z_bosons} are proportional to the
factor $t(s+t) = s(\sin \theta)^2/4$, where $\theta$ is the
scattering angle.

The SM squared amplitude for the $ \gamma Z \rightarrow l^+l^-$
collision looks like
\begin{align} \label{SM_gamma_Z_bosons}
|M_\mathrm{SM}|^{2}&= \frac{4g_Z^4}{(t - m_l^2)^2 (s + t - m_l^2 -
m_Z^2)^2}
\nonumber \\
&\times \{ -2 m_l^8[184\cos(4\theta_w) + 17\cos(8\theta_w) -
300\cos(2 \theta_w)
\nonumber \\
&- 68\cos(6\theta_w) + 195] + 4 m_l^6[ -2(59\cos(4\theta_w) +
6\cos(8\theta_w) - 94\cos(2\theta_w)
\nonumber \\
&-24\cos(6 \theta_w) + 56) m_Z^2 + 163s + 196t -8(33s + 37t)
\cos(2\theta_w)
\nonumber \\
&+ 18(9s + 10t)\cos(4\theta_w) + (15s + 16t)(\cos(8\theta_w) -
4\cos(6\theta_w))]
\nonumber \\
&+ m_l^4[(68\cos(2\theta_w) + 44\cos(6\theta_w) - 56\cos(4\theta _w)
- 11\cos(8\theta_w) - 25) m_Z^4
\nonumber \\
&+ 4[72s + 65t - 4(32s + 27t)\cos(2\theta_w) + (84s +
68t)\cos(4\theta_w)
\nonumber \\
&+ (10s + 7t)(\cos(8\theta_w) - 4\cos(6\theta_w))] m_Z^2 - 301s^2 -
436t^2 - 668st
\nonumber \\
&+ 4(125s^2 + 260ts + 156t^2)\cos(2\theta_w) - 8(3s^2 + 78st +
46t^2)\cos(4\theta_w)
\nonumber \\
&- (31s^2 + 52st + 28t^2) (\cos(8\theta_w) - 4\cos(6\theta_w))]
\nonumber \\
&+ m_l^2[ 7s^3 + 50s^2t + 92st^2 + (s^2 - 2st -
4t^2)(\cos(8\theta_w)
\nonumber \\
&- 4\cos(6\theta_w))s + 80t^3 + ( 79s^2 + 252st + 112t^2
\nonumber \\
&+ ( (5(\cos(8\theta_w) - 4\cos(6\theta_w) + 11) + 56\cos(4\theta_w)
\nonumber \\
&- 92(\cos(2\theta_w))) m_Z^2 - 141s - 226t + 4(57s +
94t)(\cos(2\theta_w))
\nonumber \\
&- 8(17s + 29t)(\cos(4\theta_w)) - 11(s + 2t)(\cos(8\theta_w) -
4\cos(6\theta_w)) ) m_Z^2
\nonumber \\
&- 4(31s^2 + 108st + 52t^2)\cos(2\theta_w) + 8(9s^2 + 34st +
17t^2)\cos(4\theta_w)
\nonumber \\
&+ (5s^2 + 28st + 16t^2) (\cos(8\theta_w) - 4\cos(6\theta_w)) )
m_Z^2
\nonumber \\
&- 4(3s^3 + 14s^2t + 24st^2 + 24t^3)\cos(2\theta_w)
\nonumber \\
&+ 8(s + 2t)(s^2 + st + 3t^2)\cos(4\theta_w) ]
\nonumber \\
&- t[ 16\cos(4\theta_w) + \cos(8\theta_w) - 28\cos(2\theta_w) -
4\cos(6\theta_w) + 19 ]
\nonumber \\
&\times (s + t - m_Z^2)(s^2 + 2t^2 + 2st - 2tm_Z^2 + m_Z^4) \} \;.
\end{align}
The contribution to the $ \gamma Z \rightarrow l^+l^-$ collision
from the gravitons $G$ is zero, since there is no $\gamma ZG$
vertex. Note that in the limit $m_l = m_Z = 0$ all squared
amplitudes depend on variables $s$ and $t(s + t) = tu$, where $u$ is
the Mandelstam variable.

For the $W^+W^- \rightarrow l^-l^+$ collision we have
\begin{align}\label{SM_W_bosons}
|M_{\mathrm{SM}}|^{2} &= -\frac{4g_{e}^{2}g_{w}^{2}}{s^{2}(s -
m_Z^2)^2} [(2s - m_Z^2)^2(\cos\theta_w)^2 (3sm_l^2 - 10tm_l^2 +
14m_l^2 m_W^2 + 5m_l^4 + 2sm_W^2
\nonumber \\
&- 10tm_W^2 + 5m_W^4 + 4s^2 + 5st + 5t^2)] + \frac{4g_{e}g_w^3}{st(s
- m_Z^2)}[(2s - m_Z^2)\cos \theta_w
\nonumber \\
&\times (sm_l^2 + tm_l^2 + m_{l}^{2}m_{w}^{2} - 2m_l^4 - sm_W^2 -
2tm_W^2 + m_W^4 + 3st + t^2)]
\nonumber \\
&- \frac{4g_w^4}{t^{2}} [m_l^4 - 2m_l^2m_W^2 -2tm_W^2 + m_W^4 + st +
t^2] \;,
\end{align}
where $g_w=g_e/\sin\theta_w$, and a neutrino mass is taken to be
zero.

The interference term looks like
\begin{align}\label{int_W_bosons}
|M_{\mathrm{int}}|^2 & =[S(s) + S^{\star}(s)] \Big[\frac{g_eg_w(2s -
m_{Z}^2)\cos \theta _w}{2s(s - m_Z^2)^2}(-2m_l^2 - 2m_W^2 + s + 2t)
\nonumber \\
& \times (sm_l^2 - 4tm_l^2 + 4m_l^2m_W^2 + 2m_l^4 +sm_W^2 - 4tm_W^2
+ 2m_W^4 + s^2 + 2st + 2t^2)
\nonumber \\
& -\frac{g_w^2}{24t} (3s^{2}m_l^2 - 6sm_l^2m_W^2 + 3sm_l^4 +
10tm_l^2m_W^2 - 18tm_l^4 + 4m_l^4m_W^2
\nonumber \\
&- 10m_l^2m_W^4 + 12m_l^6 - 3s^2m_W^2 - 18stm_W^2 + 3sm_W^4 -
18t^2m_W^2 + 18tm_W^4)
\nonumber \\
&- 6m_W^6 + 9s^2t + 15st^2 + 6t^3 \Big] .
\end{align}

Finally, the squared amplitude $|M_{\mathrm{KK}}(W^+W^- \rightarrow
l^-l^+)|^{2}$ is obtained from eq.~\eqref{KK_Z_bosons} by the
replacement $m_Z \rightarrow m_W$.



\setcounter{equation}{0}
\renewcommand{\theequation}{D.\arabic{equation}}

\section*{Appendix D. Squared amplitudes for $l^+l^-\rightarrow l^+l^-$ scattering}
\label{app:D}

Here we present the result of our calculations of the squared
amplitudes for the $l^+l^-\rightarrow l^+l^-$ process (both incoming
and outgoing leptons have the same flavor). The SM squared amplitude
has both the $Z$ boson and photon contributions. The latter one is
given by the formula
\begin{align}\label{M2_SM}
|M_{\mathrm{SM}}|^2&= \frac{16e^{4}}{s^2 t^2}[m_l^4 (5s^2 + 11st +
5t^2) + 4m_l^2(s + t)(s^2 + 5st + t^2)
\nonumber \\
&+(s^2 + st + t^2)^2] \;,
\end{align}
where $m_l$ is the lepton mass. Note that the photon contribution to
$|M_{\mathrm{SM}}|^2$ is dominant. That is why, we do not present
(rather complicated) analytical expression for the $Z$ boson
contribution to $|M_{\mathrm{SM}}|^2$. The graviton squared
amplitude is defined by KK graviton exchanges in $s$- and
$t$-channels,
\begin{align}\label{M2_KK}
|M_{\mathrm{KK}}|^2&= \frac{1}{4608} \{ |S(s)|^2 F_1(s,t) + |S(t)|^2
F_1(t,s)
\nonumber \\
&+ [S(s)S(t)^{\star} + S(s)^{\star}S(t)]F_2(s,t) \} \;.
\end{align}
Finally, the interference term of $|M|^2$ is equal to (neglecting
small $Z$ boson contribution)
\begin{align}\label{M2_int}
|M_{\mathrm{int}}|^2&= -\frac{e^{2}}{24st}\{ [S(s) +
S(t)^{\star}]F_3(s,t) + [S(s)^{\star} + S(t)]F_3(t,s) \} \;.
\end{align}
Here $S(s)$ is defined by eqs.~\eqref{S}, \eqref{A_epsilon}, and the
following functions are introduced
\begin{align}\label{F1}
F_1(s,t)&= 6656m_l^8 + m_l^6(8576s + 10752t) + m_l^4(3440s^2 +
7296t^2 + 10752st)
\nonumber \\
&+ m_l^2(360s^3 + 2376s^2t + 4320st^2 + 2304t^3)
\nonumber \\
&+ 9s^4 + 90s^3t + 378s^2t^2 + 576st^3 + 288t^4 \;,
\end{align}
\begin{align}\label{F2}
F_2(s,t)&= 7552m_l^8 + 15168m_l^6(s + t) + m_l^4(7248(s^2 +
t^2)+14968st)
\nonumber \\
&+ m_l^2(1032(s^3 + t^3)+3690st(s + t))
\nonumber \\
&+ 36(s^4 + t^4) + 225st(s + t) + 378s^2t^2 \;,
\end{align}
\begin{align}\label{F3}
F_3(s,t)&= m_l^6(576s + 512t) + m_l^4(552s^2 + 448t^2 + 1120st)
\nonumber \\
&+ m_l^2(86s^3 + 360s^2t + 432st^2 + 144t^3)
\nonumber \\
&+ 3s^4 + 21s^3t + 45s^2t^2 + 48st^3 + 24t^4 \;,
\end{align}
where $m_l$ is the lepton mass. Note that $F_2(s,t) = F_2(t,s)$.




\end{document}